\documentclass[aps,prb,twocolumn,superscriptaddress]{revtex4}
\usepackage{color}
\usepackage{booktabs}
\usepackage{epsfig}
\usepackage{graphicx}     
\usepackage[dvipdfmx]{hyperref}
\hypersetup{colorlinks=true,citecolor=blue,linkcolor=red}
\usepackage[all]{hypcap}
\usepackage{url}
\begin{document}

\title{ Local-fields and disorder effects in free-standing and embedded Si nanocrystallites }

\author{\firstname{Roberto} \surname{Guerra}}
\affiliation{Centro S3, CNR-Istituto di Nanoscienze, via Campi 213/A I-41100 Modena Italy.}
\affiliation{Dipartimento di Fisica, Universit\`a di Modena e Reggio Emilia, via Campi 213/A I-41100 Modena Italy.}

\author{\firstname{Elena} \surname{Degoli}}
\affiliation{Centro S3, CNR-Istituto di Nanoscienze, via Campi 213/A I-41100 Modena Italy.}
\affiliation{Dipartimento di Scienze e Metodi dell'Ingegneria, Universit\`a di Modena e Reggio Emilia, via Amendola 2 Pad. Morselli I-42100 Reggio Emilia Italy.}

\author{\firstname{Margherita} \surname{Marsili}}
\author{\firstname{Olivia} \surname{Pulci}}
\affiliation{European Theoretical Spectroscopy Facility (ETSF) and CNR-INFM, Dept. of Physics, Universit\'a di Roma "Tor Vergata" Via della Ricerca Scientifica 1, I-00133 Roma, Italy.}

\author{\firstname{Stefano} \surname{Ossicini}}
\affiliation{Centro S3, CNR-Istituto di Nanoscienze, via Campi 213/A I-41100 Modena Italy.}
\affiliation{Dipartimento di Scienze e Metodi dell'Ingegneria, Universit\`a di Modena e Reggio Emilia, via Amendola 2 Pad. Morselli I-42100 Reggio Emilia Italy.}

\begin{abstract}
The case study of a 32-atoms Si nanocrystallite (NC) embedded in a SiO$_2$ matrix, both crystalline and amorphous, or free-standing with different conditions of passivation and strain is analyzed through ab-initio approaches. The Si$_{32}$/SiO$_2$ heterojunction shows a type I band offset highlighting a separation between the NC plus the interface and the matrix around. The consequence of this separation is the possibility to correctly reproduce the low energy electronic and optical properties of the composed system simply studying the suspended NC plus interface oxygens with the appropriate strain. Moreover, through the definition of an optical absorption threshold we found that, beside the quantum confinement trend, the amorphization introduces an additional redshift that increases with increasing NC size: i.e. the gap tends faster to the bulk limit. Finally, the important changes in the calculated DFT-RPA optical spectra upon inclusion of local fields point towards the need of  a proper treatment of the optical response of the interface region.
\end{abstract}
\pacs{78.20.Ci; 78.40.Fy; 78.55.Ap}
\maketitle

\section{Introduction}
Light emission in the visible spectral range and at room temperature has been observed in the last two decades in porous Si before and in nanostructured Si after. Si nanocrystallites embedded in a silicon dioxide matrix, in particular, have attracted much attention in recent years \cite{1,2,3}. In these structures, Si acts as the low-energy gap material. The strong quantum confinement induced by the dielectric matrix enlarges the band-gap and increases the spatial overlap of the electron and hole wave functions, thus increasing the spontaneous emission rate and shifting the emission peak to energies higher than bulk Si band-gap. Despite the large number of experiments and theoretical studies, many question are still open concerning the nature of the optical properties of Si-NCs embedded in a silica matrix. Actually, too many parameters, like size, shape, strain and interface structure, influence the spectral response of the system. The characterization of these quantities is thus of fundamental importance in order to determine the best condition for light emission and optical gain.
\\In this work, we present DFT calculations on a crystalline and amorphous 32-atoms Si-NC, passivated with hydrogens, OH groups, and embedded in a SiO$_2$ matrix. The comparison between the embedded and suspended NCs can give insights on the role played by strain in the determination of the electronic and optical properties of the composed system, while the different passivation regimes can reveals the contribution given separately by the NC, the interface and the matrix to the properties of the whole system.
\\The paper is organized as follows. A description of the theoretical methods and of the systems is given in section \ref{theory}. The electronic and optical properties of the considered systems are analyzed in section \ref{ele-opt} while conclusions are presented in section \ref{concl}.

\section{Theoretical method}\label{theory}

\begin{figure*}[t!]
  \centering
  \begin{minipage}[c]{0.7\textwidth}
    \centerline{\large{a)}~~\includegraphics*[width=5cm]{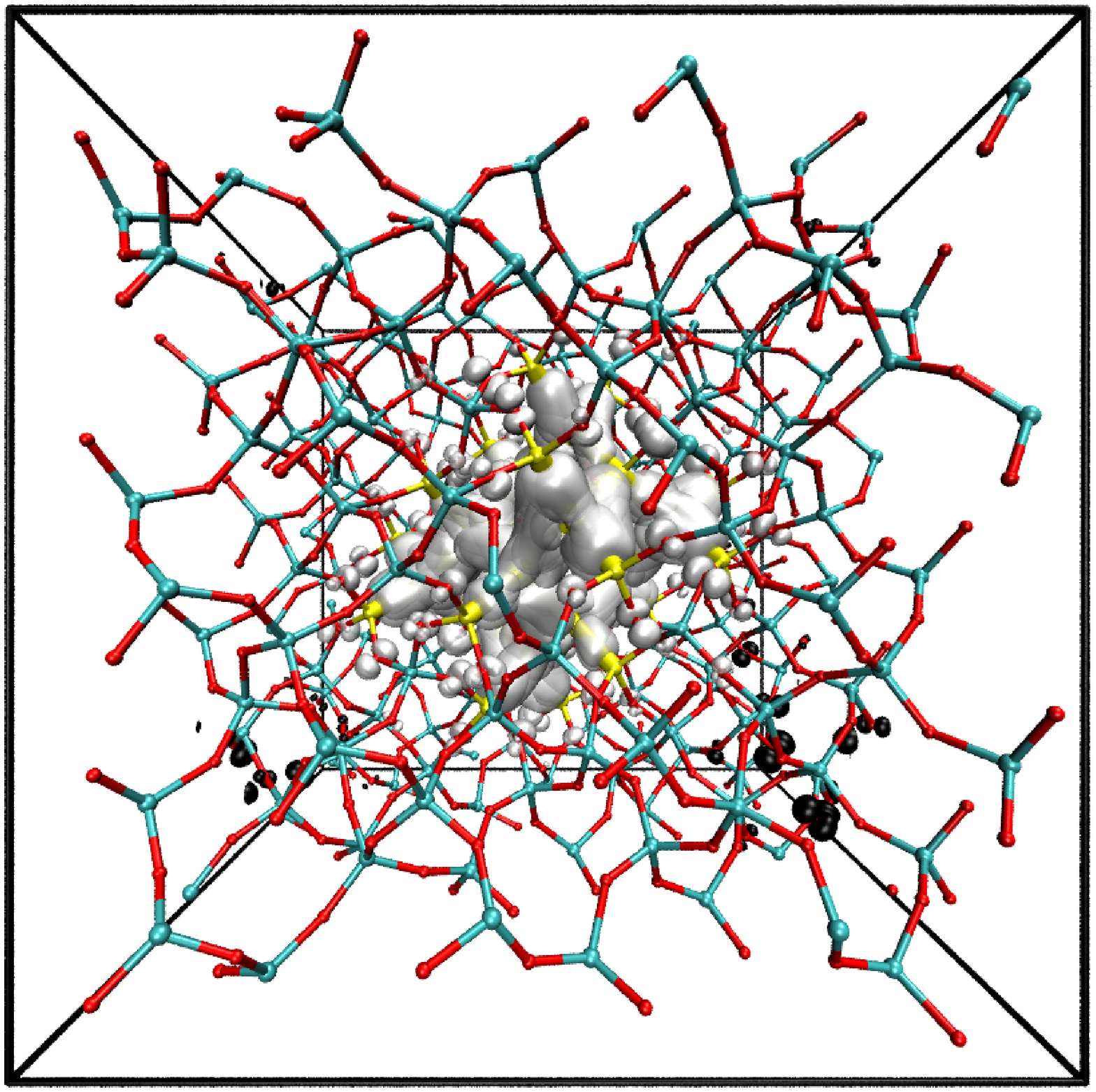}~~~~
                \large{b)}~~\includegraphics*[width=5cm]{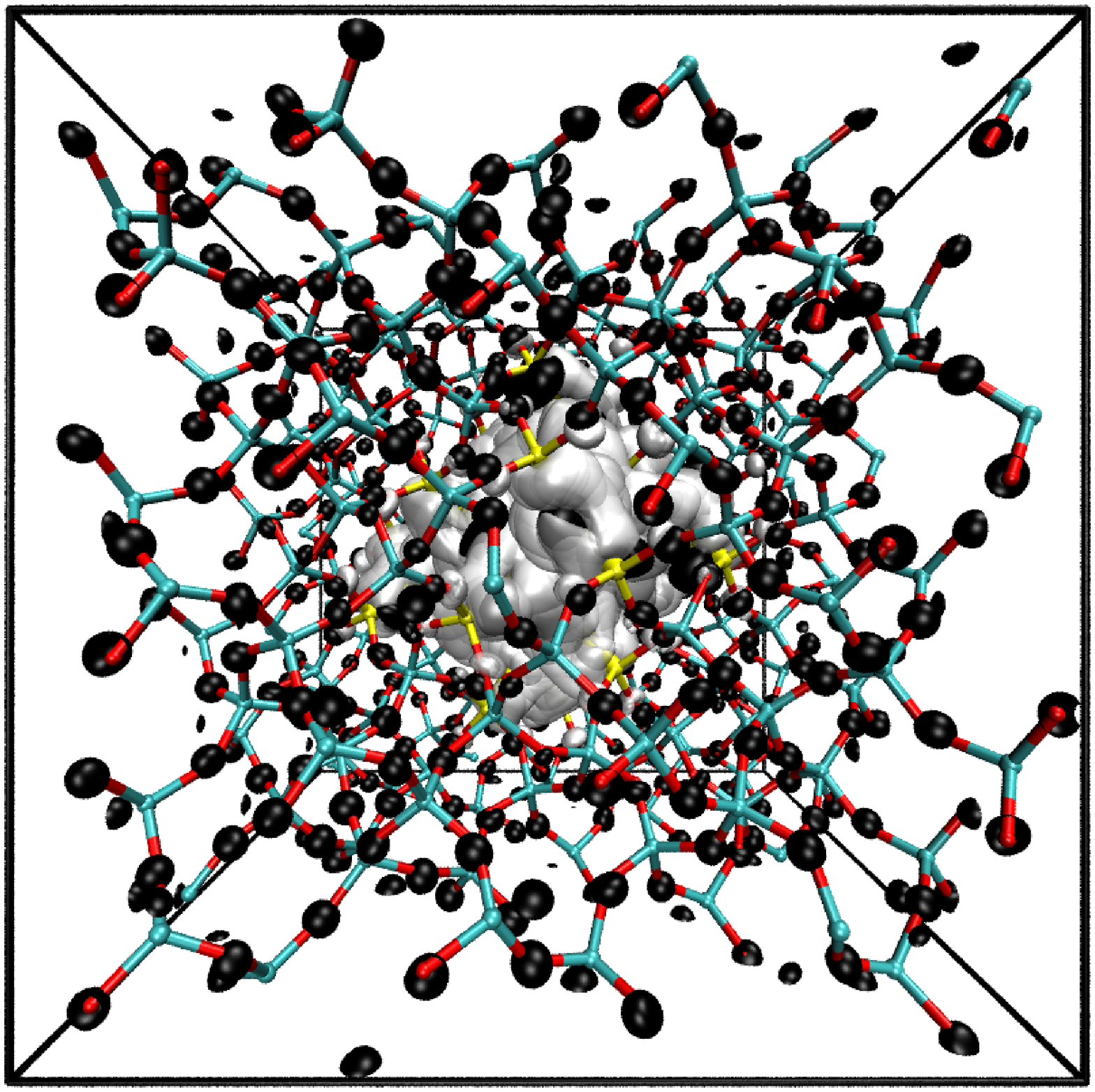}}
    \centerline{\large{c)}~~\includegraphics*[width=5cm]{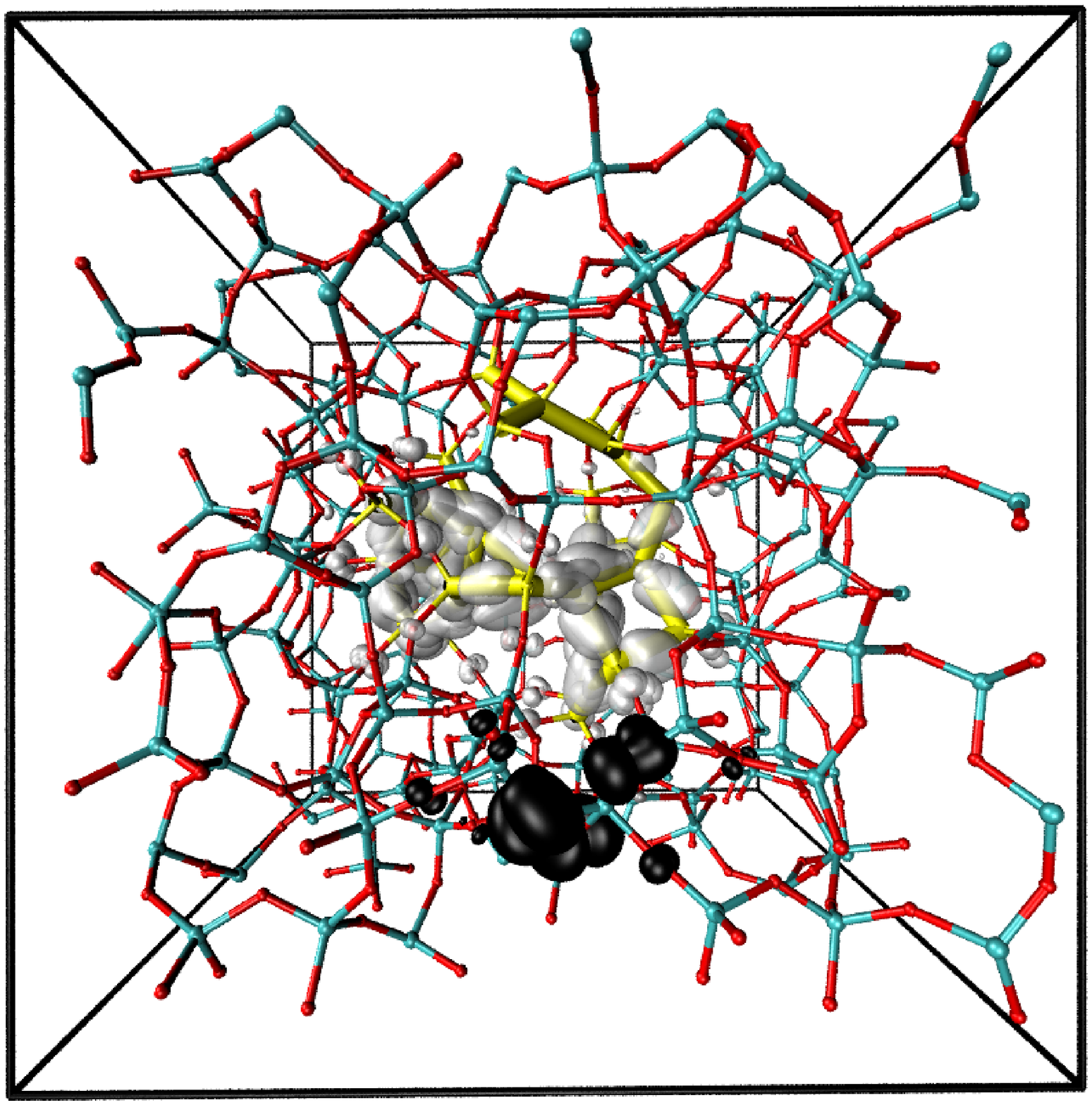}~~~~
                \large{d)}~~\includegraphics*[width=5cm]{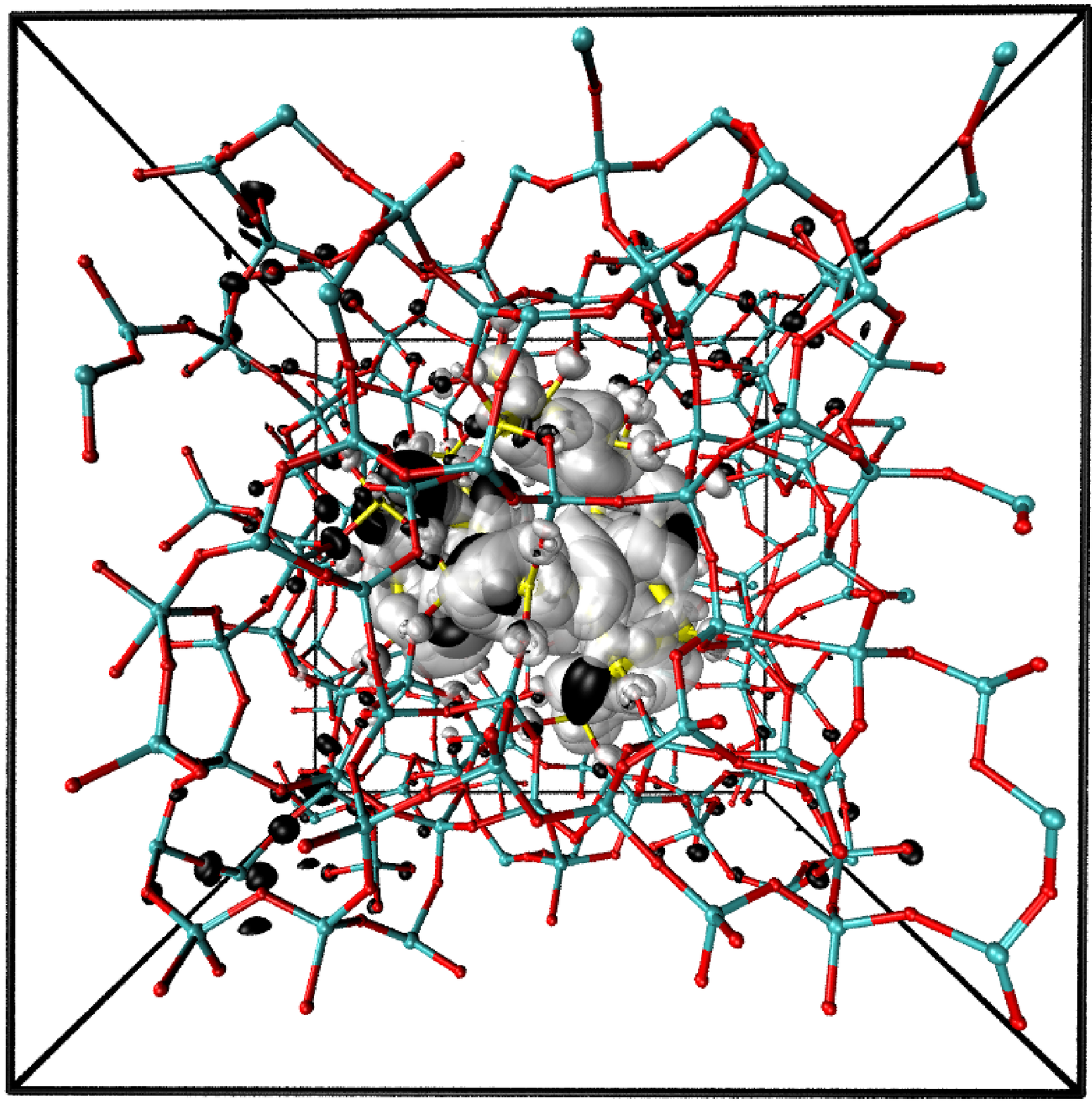}}
  \end{minipage}
  \begin{minipage}[c]{0.25\textwidth}
    \centering
    \caption{Stick and ball pictures of the final optimized structure of Si$_{32}$ in $\beta$-cristobalite matrix (top panels) or in amorphous glass (a-SiO$_2$)(bottom panels). Red (dark gray) spheres represent the O atoms, cyan (gray) the Si of the matrix, and yellow (light gray) the Si atoms of the NC. On the structures, Kohn-Sham orbitals at 10\% of their maximum amplitude are reported. In white are the sum of all band-edges orbitals localized on the NC while in black is first the state delocalized on the matrix: a) NC: from HOMO to HOMO-15, matrix: HOMO-16, b) NC: from LUMO to LUMO+17, matrix: LUMO+18, c) NC: from HOMO to HOMO-8, matrix: HOMO-9, d) NC: from LUMO to LUMO+18, matrix: LUMO+19.}\label{HL}
  \end{minipage}
\end{figure*}

{\it Ab-initio} calculations of  structural, electronic and optical properties for a 32 atoms Si-NC with different surface passivations or embedded in a SiO$_2$ matrix have been carried out within Density Functional Theory (DFT) \cite{DFT}. In the case of the embedding medium we have also analysed the role played by disorder considering a crystalline phase or an amorphous glass. For the crystalline structure we have considered a cubic cell (of side L$\simeq 2.1$ nm) of $\beta$-cristobalite (BC) SiO$_2$ which is well known to give rise to one of the simplest Si/SiO$_2$ interface because of its diamond-like structure \cite{BC}. The NC/matrix crystalline structure has been obtained from a Si$_{216}$O$_{432}$ matrix by removing the 36 oxygens included in a sphere of radius 0.6 nm ~placed at the center of the cubic supercell (see Fig.\ref{HL}, top panel). The result is a structure of 216 Si and 396 O atoms with 32 Si bonded together to form a small crystalline skeleton with T$_d$ local symmetry before relaxation. In such core, Si atoms show a larger bond length (3.1 \AA) with respect to that of the  Si bulk structure (2.35 \AA). No defects (dangling bonds) are present at the interface and all the O atoms at the NC surface are single bonded with the Si atoms of the cluster. The dimension of the  matrix preserves a separation of at least 1 nm between the NCs replica, ensuring a proper shielding of the introduced strain \cite{mluppi1,mluppi2,flyura}. The optimized structure has been achieved by fully relaxing the atomic positions and the cell shape and size. The host matrix gets distorted after the relaxation (essentially due to the metastable nature of the BC), producing an optical behaviour similar to that of a completely amorphized silica, but with practically irrelevant effects on the NC cristallinity. Therefore, the final system should be considered more like a crystalline NC in an amorphous SiO$_2$.
\\Together with the crystalline structure, the complementary case of an amorphous silica (a-SiO$_2$) has been considered. The glass model has been generated using classical molecular dynamics (MD) simulations of quenching from a melt using semi-empirical ionic potentials \cite{Feuston}. The amorphous dot structure has been obtained starting from a Si$_{216}$O$_{432}$ amorphous silica cell by removing the 40 oxygen atoms included in a sphere of radius  0.6 nm placed at the center of the cell, as shown in Fig. \ref{HL} (bottom panel). Calculations of both the crystalline and amorphous structures have been performed using the SIESTA code \cite{siesta1,siesta2} with core-corrected Troullier-Martins pseudopotentials and a cutoff of $150 Ry$ on density. No additional external pressure or stress were applied. Atomic positions and cell parameters have been left totally free to move. We note that the crystalline and the amorphous clusters present a nearest-neighbor distance of about 2.46 \AA, strained respect to the typical Si-bulk value (2.35 \AA), in good agreement with the outcomes of Yilmaz et al. \cite{yilmaz}.
\\In parallel to the crystalline and amorphous Si$_{32}$/SiO$_2$ systems, three other structures have been studied: (i) the pure matrix (in the same phases as used for the embedded NC calculations), (ii) the isolated NCs as extracted from the relaxed NC-silica complexes and capped by hydrogen atoms (Si$_{32}$-H), and (iii) the NCs together with the first interface oxygens extracted as in point (ii) and then passivated by hydrogen atoms (Si$_{32}$-OH). In the last two cases only the hydrogen atoms have been relaxed. The goal is to distinguish between the properties that depend only on the embedded NC from those that are instead influenced by the presence of the matrix. The comparison of the results relative to different passivation regimes (H or OH groups) could give some insight on the role played by the interface region. As last point, the structures obtained as in point (ii) and (iii) have been fully relaxed in order to infer, through the comparison with the unrelaxed structures, the role played the strain induced by the lattice mismatch between Si and SiO$_2$.
\\Electronic and optical properties of the relaxed structures have been obtained in the framework of DFT, using the ESPRESSO package\cite{espresso}. Calculations have been performed using norm-conserving pseudopotentials within the LDA approximation with a Ceperley-Alder exchange-correlation potential, as parametrized by Perdew-Zunger. An energy cutoff of 60 Ry on the plane wave basis have been considered.
\\Local fields (LF) contribution to the optical properties in the random phase approximation (RPA) has also been computed using the EXC code\cite{exc_code}.

\begin{figure}[t!]
  \begin{center}
  \includegraphics*[width=8cm]{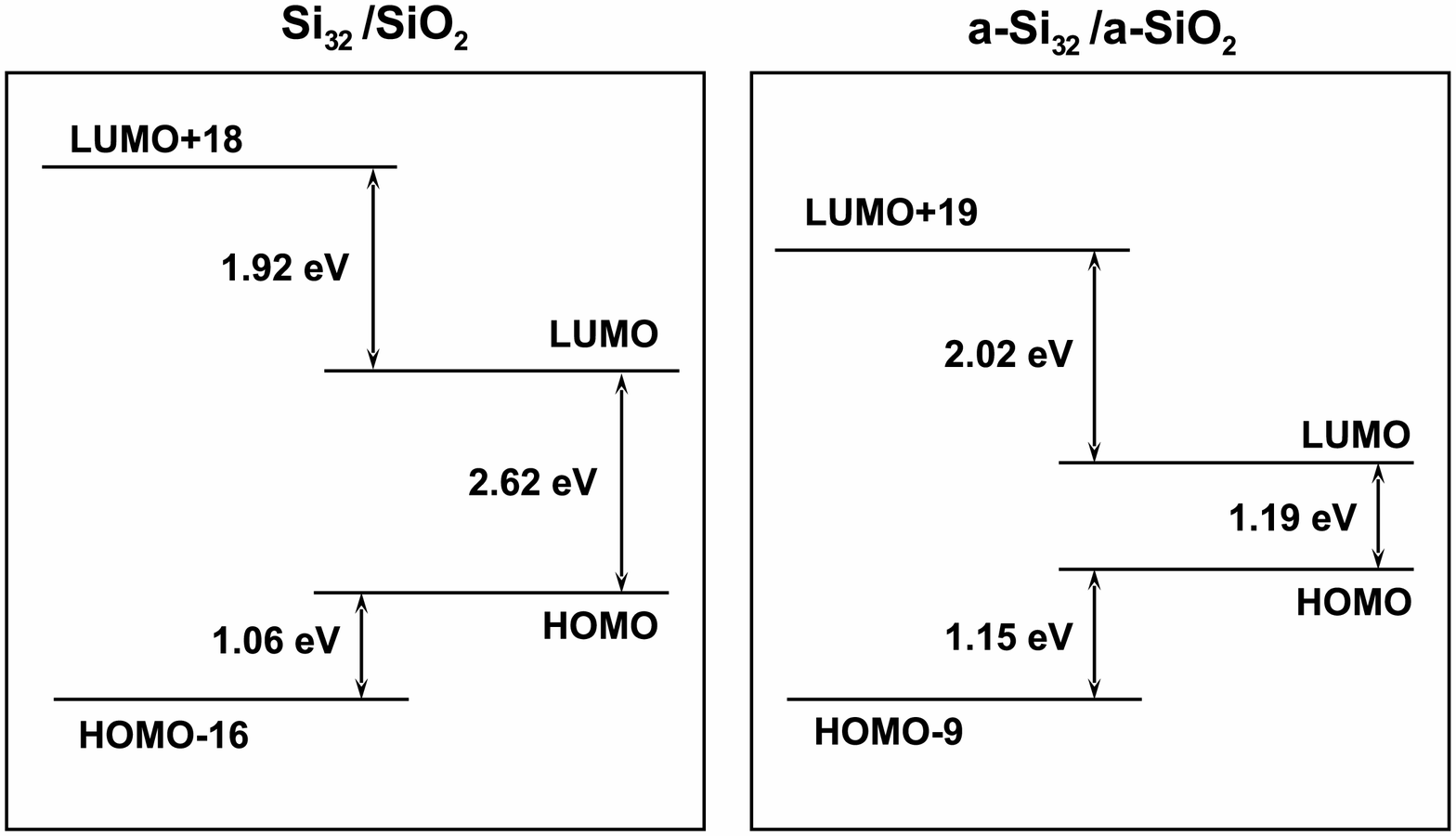}
  \caption{\label{offset} Valence and conduction band offset calculated for the Si$_{32}$ NC in BC (left panel) and in the amorphous glass (right panel).}
  \end{center}
\end{figure}
\begin{figure*}[t!]
  \centering
  \begin{minipage}[c]{0.7\textwidth}
  \centerline{\large{a)}\hspace*{-1cm}\includegraphics*[width=4.5cm,angle=270]{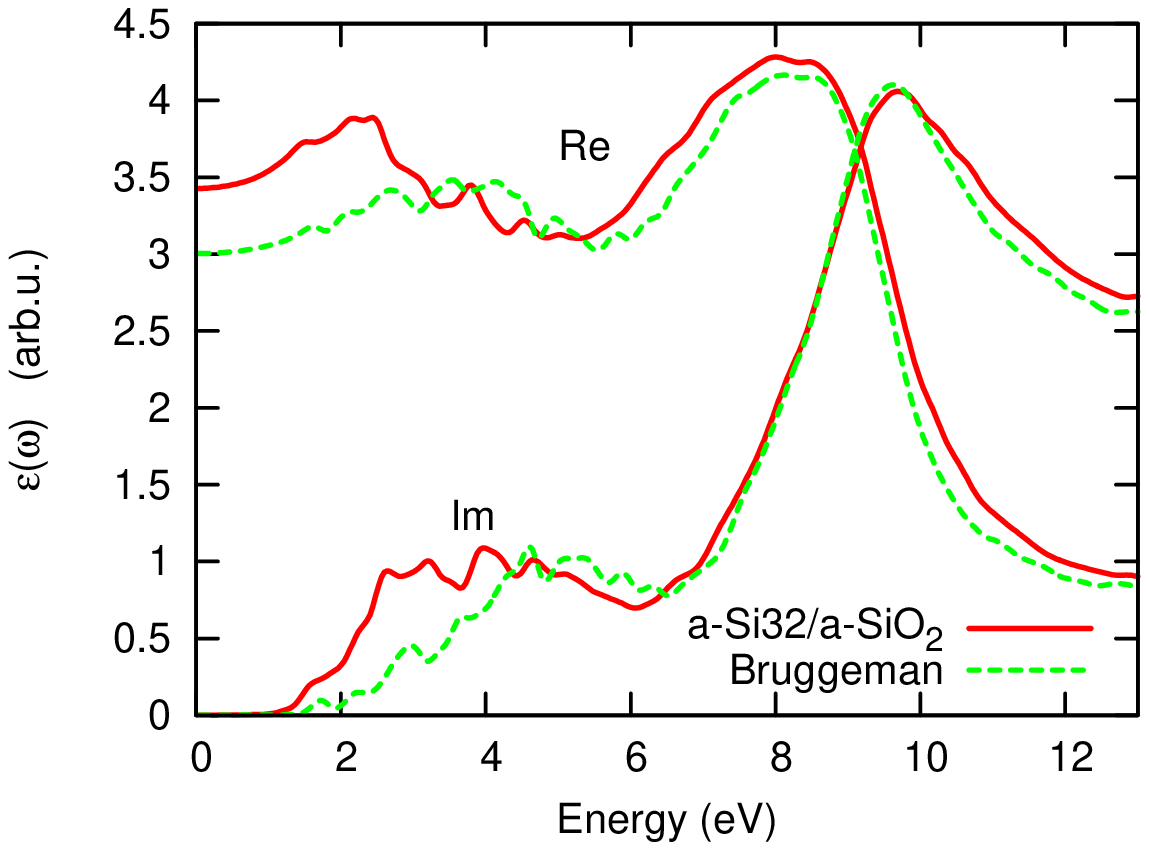}~~~~
              \large{b)}\hspace*{-1cm}\includegraphics*[width=4.5cm,angle=270]{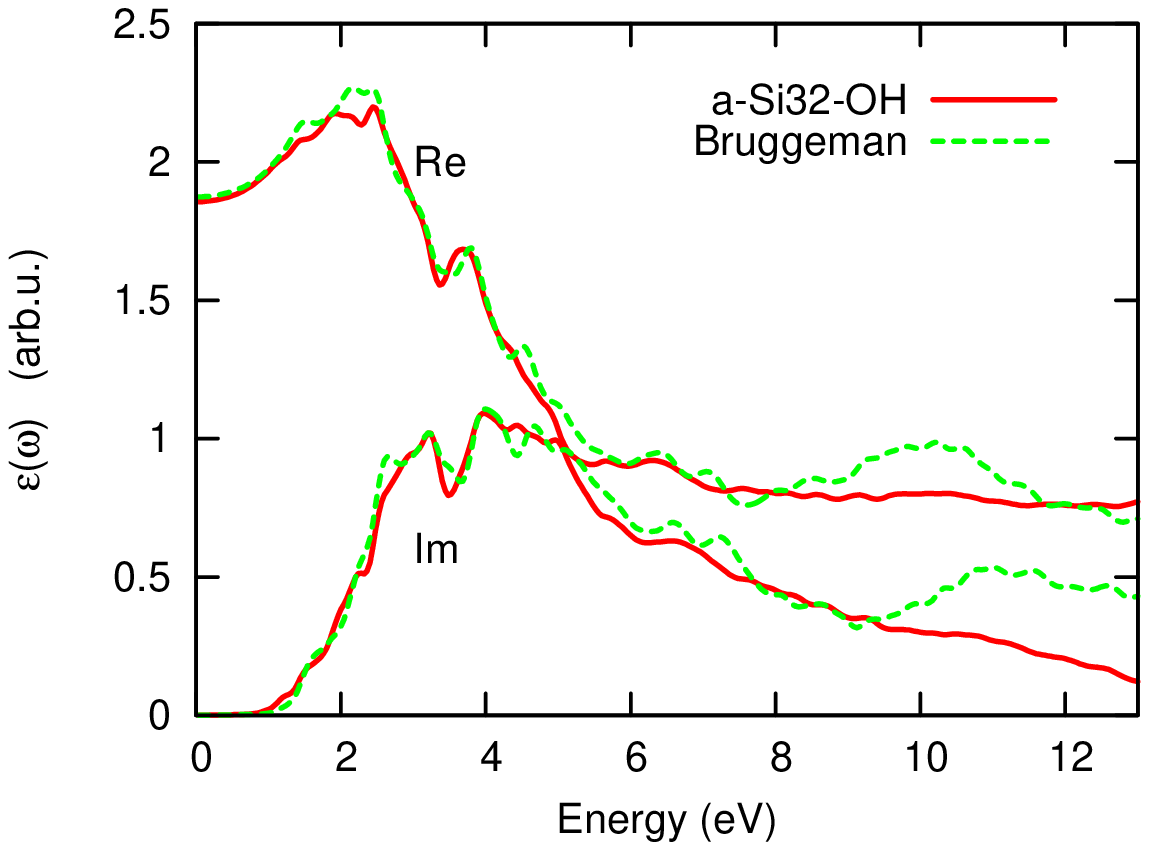}}
  \centerline{\large{c)}\hspace*{-1cm}\includegraphics*[width=4.5cm,angle=270]{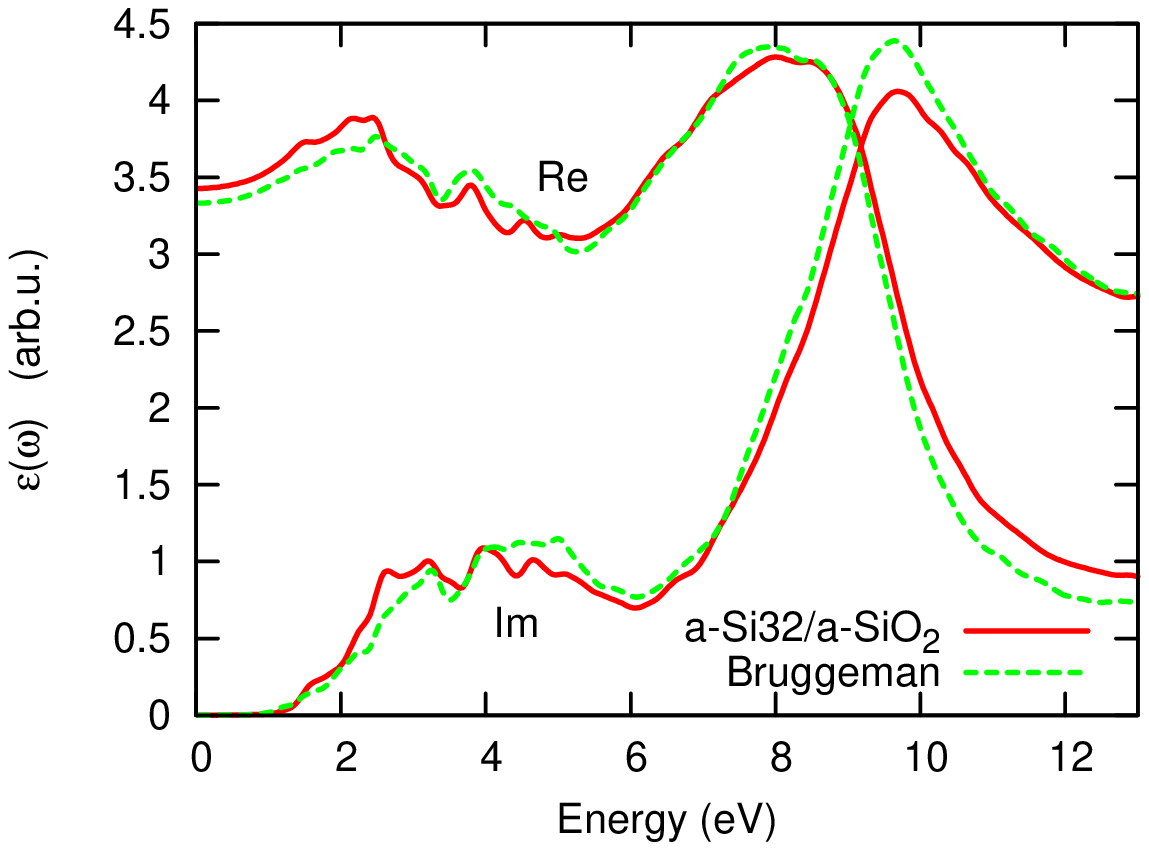}~~~~
              \large{d)}\hspace*{-1cm}\includegraphics*[width=4.5cm,angle=270]{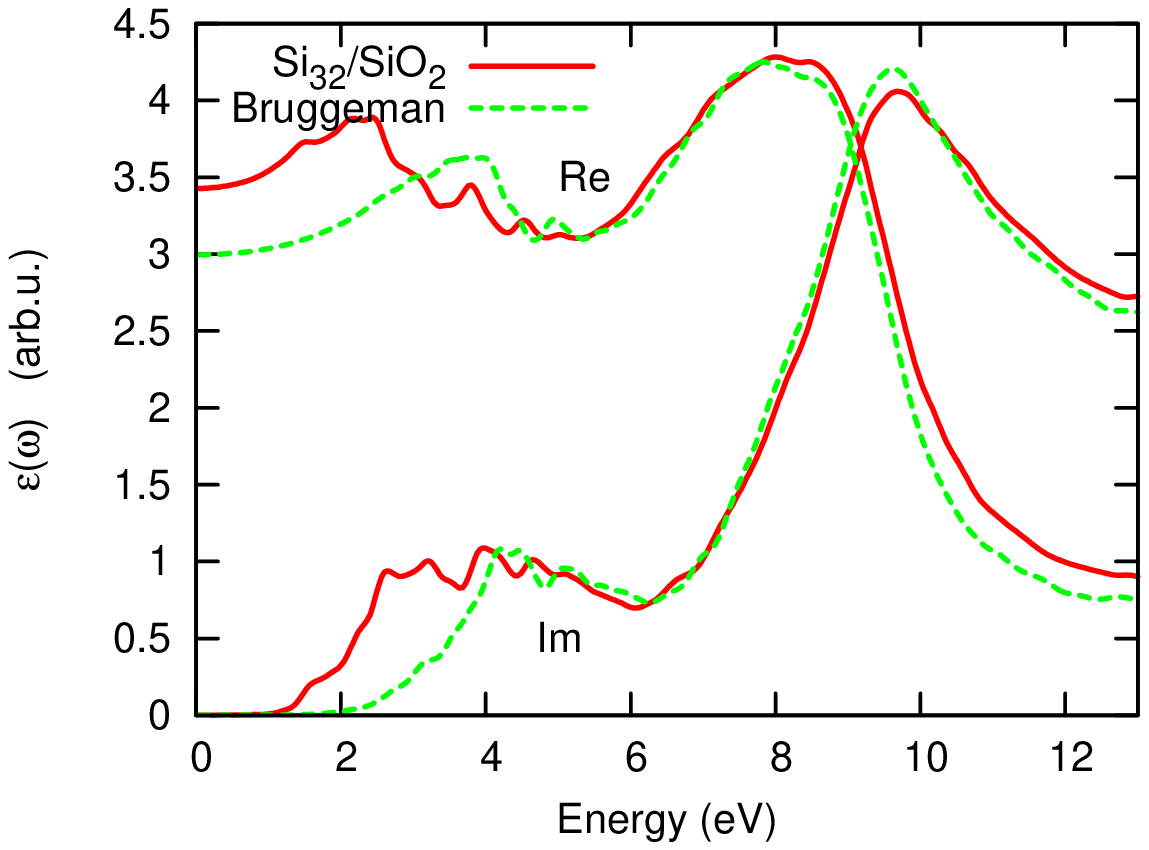}}
  \end{minipage}
  \begin{minipage}[c]{0.25\textwidth}
  \centering
  \caption{\label{BEMA} Comparison between the dielectric function of a) the a-Si$_{32}$/a-SiO$_2$ and that obtained by the BEMA composing the strained a-Si$_{32}$-H NCs and the a-SiO$_2$, b) the strained a-Si$_{32}$-OH and that obtained by the inverse-BEMA composing the a-Si$_{32}$/a-SiO$_2$ and the a-SiO$_2$, c) the a-Si$_{32}$/a-SiO$_2$ and that obtained by the BEMA composing the strained a-Si$_{32}$-OH NCs and the a-SiO$_2$, d) the a-Si$_{32}$/a-SiO$_2$ and that obtained by the BEMA composing the relaxed a-Si$_{32}$-OH NCs and the a-SiO$_2$. Real and imaginary parts are indicated by labels.}
  \end{minipage}
\end{figure*}

\section{The electronic and optical properties}\label{ele-opt}
When an interface is created between two different semiconductors, as in our case between Si and SiO$_2$, a valence and conduction band offset takes place. Depending on the relative position of the band edges states in the two material one can have type I (straddling gap) or type II (staggered gap) heterojunction. Through the energy band character we have roughly evaluated the conduction and valence band offset for our Si$_{32}$ NC embedded both in the crystalline and in the amorphous silica (see Fig.\ref{offset}). Looking at energy bands localization we have found that all the band edges states from HOMO to HOMO-15 and from LUMO to LUMO+17 in BC (Fig. \ref{HL}a and \ref{HL}b, white orbitals) and from HOMO to HOMO-8 and LUMO to LUMO+18 in a-SiO$_2$ (Fig. \ref{HL}c and \ref{HL}d, white orbitals) are fully localized on the NC. The states HOMO-16 and LUMO+18 in BC (Fig. \ref{HL}a and \ref{HL}b, black orbitals) and HOMO-9 in silica (Fig. \ref{HL}c, black orbitals) are the first states fully localized on the matrix, while LUMO+19 in silica (Fig. \ref{HL}d, black orbital) is localized on both the matrix and the NC. No pure states on the matrix exists at least for 5 eV above the gap; this a consequence of the amorphous nature of the matrix. Despite the uncertainty on the silica conduction band offset this configuration is typical of a type I heterojunction as reported in Fig. \ref{offset}. The energy values reported in Fig. \ref{offset} are underestimated being DFT results. GW corrections are of the order of 5 eV for silica \cite{Chang} and of about 1 eV for NC \cite{PRB79}.
\\The electronic properties just described determine the optical response of the system: we expect to have a low energy range spectrum dominated by the NC while the matrix control the higher energy optical transitions. What about the interface region? The application of the Bruggeman Effective Medium Approximation (BEMA) \cite{bruggeman} on Si-NCs is the natural choice to investigate the interplay between the dielectric function of the NC, the SiO$_2$ matrix, and the composite system. Also, the applicability of the BEMA constitutes a solid tool to test the level of interaction between the embedded NC and the embedding matrix. Initially, through the BEMA we have calculated the dielectric function of the composite system, $\varepsilon$, by combining that of the host matrix, $\varepsilon_h$, and that of the NC, $\varepsilon_c$. This procedure is not well defined when applied to the crystalline NC because BC matrix loses its crystallinity after the creation of the NC, and therefore the BC is not representative of the final embedding medium. Therefore in all cases we should use the amorphous silica for a better application of the BEMA. The real parts of the dielectric functions are calculated from the corresponding DFT-RPA imaginary parts, $\varepsilon_2$, through the Kramers-Kronig relations.  The comparison of the total dielectric functions obtained starting from the $\varepsilon_2$ of the pure (idrogenated) NC and that of the pure silica glass, with that of the a-Si$_{32}$ embedded NCs, is reported in Fig. \ref{BEMA}a. We note a clear mismatch between the curves, especially in the low-energy range. Actually, the BEMA is based on the assumptions that the embedded NC is spheric and not interacting with the SiO$_2$. The first assumption is essentially satisfied by construction, because we built the NC using a spherical cut-off. Then, the second assumption should be conclusive for the mismatch.
\begin{table*}[t!]
  \begin{tabular}{p{35mm}p{22mm}p{22mm}p{22mm}p{22mm}p{22mm}} \hline
                & SiO$_2$ & Si$_{2}$/SiO$_{2}$ & Si$_{5}$/SiO$_{2}$& Si$_{10}$/SiO$_{2}$ &Si$_{32}$/SiO$_{2}$\\ \hline
    Crystalline &  6.72 &  5.66  & 4.20 & 3.06 &2.46  \\ \hline
    Amorphous   & 6.25  & 5.49   & 3.76 & 1.91 &1.10  \\ \hline
    $\Delta$    & 0.47  & 0.17   & 0.44 & 1.15 &1.36 \\ \hline
  \end{tabular}
  \caption{Calculated $E_g^o$) values (in eV) for different size Si nanocrystals embedded in crystalline or amorphous silica, compared with that of pure SiO$_2$. In the last line the difference between the crystalline and amorphous case are reported.}\label{optgap}
\end{table*}
\\In order to find the dielectric function of the non-interacting system, we adopt the alternative approach of the BEMA (inverse-BEMA), in which we obtain the $\varepsilon_c$ of the NC from the difference between the DFT-calculated $\varepsilon$ of the composite a-Si$_{32}$/a-SiO$_2$ system and $\varepsilon_h$ of the pure silica. The result is shown in Fig. \ref{BEMA}b. We note that the dielectric function of the non-interacting system matches that of the strained a-Si$_{32}$-OH (the Si NC plus the first shell oxygen atoms around it as extracted by the matrix, i.e. the Si NC plus the interface region), especially in the 0-6 eV energy range. This strongly supports the idea that a true separation is established between the NC+interface system and the remaining silica matrix. This separation, due to the non-interacting behaviour of the parts, allows an excellent description of the absorption at low energies. If, instead, through the BEMA, we compose the dielectric function of the strained a-Si$_{32}$-OH NCs and that of the a-SiO$_2$ we obtain the result shown in Fig. \ref{BEMA}c where, as expected, a good agreement with the DFT-calculated $\varepsilon$ of the embedded system is obtained.
\\To underline the important role played by the strain that the matrix induce on the NC, in Fig. \ref{BEMA}d, we compare the $\varepsilon_2$ obtained through the BEMA combining the contributions of the relaxed a-Si$_{32}$-OH NC and of the pure silica glass, with that of the a-Si$_{32}$ embedded NC. The result is definitely worse with respect to Fig. \ref{BEMA}c where strain was included.
\\From the DFT-RPA optical spectra it is possible to introduce a new quantity, $E_g^o$ , which is defined as the optical absorption threshold when all the transitions with an intensity lower than 1\% of the highest peak are neglected. By neglecting very low oscillator strength optical transitions we introduce a sort of ``instrument resolution'' that can connect theoretical and experimental results concerning the optical activity of a system. The obtained $E_g^o$ values for the embedded systems are reported in Table \ref{optgap}. Here $E_g^o$ has been calculated for Si-NC containing 2, 5, 10, and 32 atoms and compared with that of pure silica. While the increasing of the NC size always results in a reduction in $E_g^o$, the effect of the amorphization is to introduce an additional redshift ($\Delta$) that progressively increase and becomes relevant for larger NCs. As a consequence the band gap tend faster to the bulk limit in the amorphous system with respect to the crystalline one.
\begin{figure}[ht]
  \includegraphics*[width=4.5cm,angle=270]{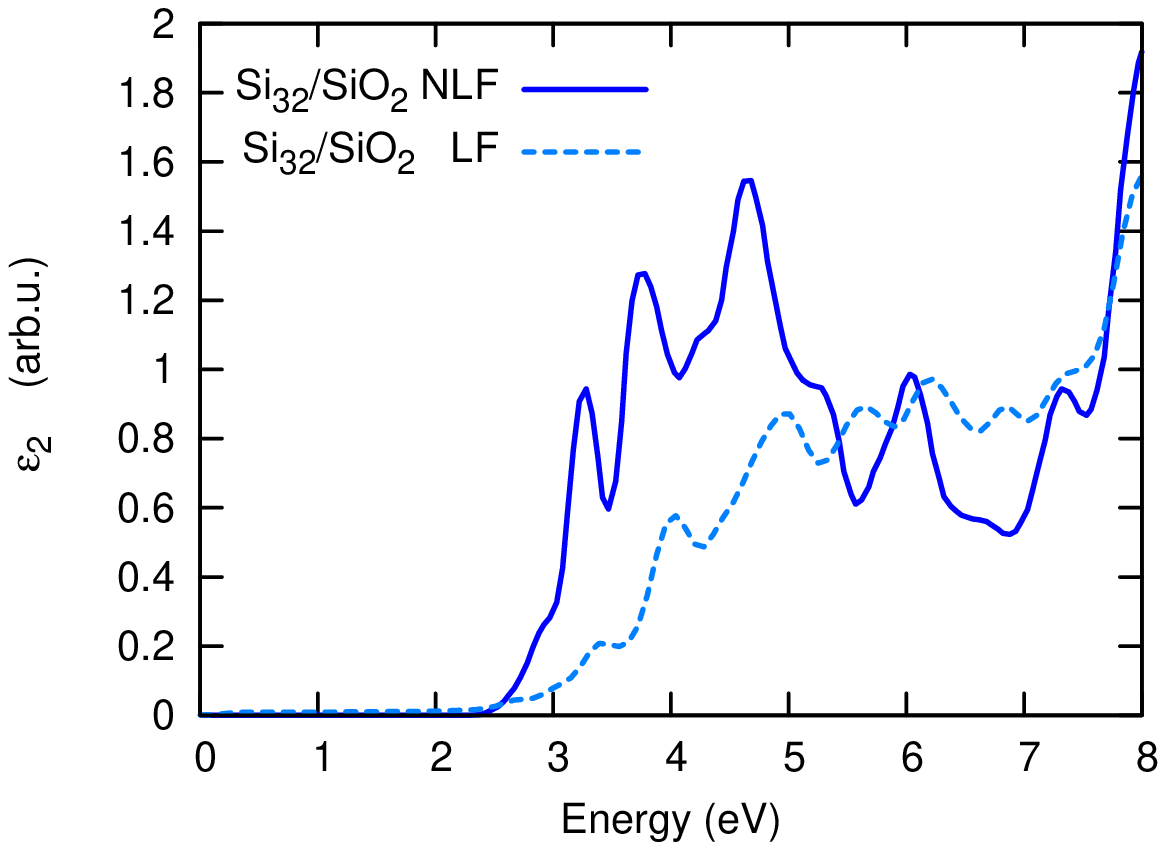}
  \includegraphics*[width=4.5cm,angle=270]{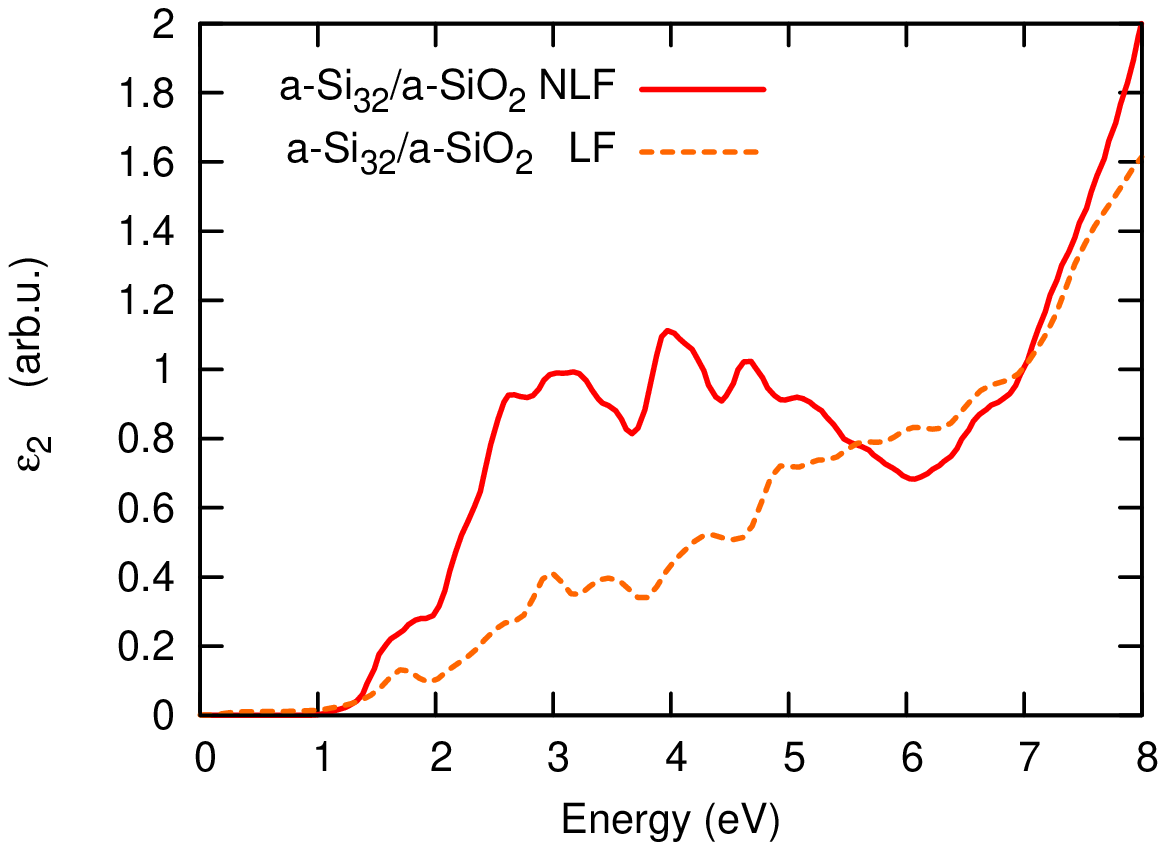}
  \caption{\label{LFvsNLF} DFT-RPA calculated imaginary part of the dielectric function for crystalline (top) and amorphous (bottom) Si$_{32}$ embedded NCs, with and without LF contribution.}
\end{figure}
Finally, we investigated the  optical properties of the Si$_{32}$ and a-Si$_{32}$ embedded NCs. In accordance with Delerue and coworkers \cite{Delerue} that found, for Si-NC larger than 1.2 nm (this is approximatively our cluster dimension), a mutual cancellation of self-energy and Coulomb corrections, we neglect Many-Body effects. Local fields effects, instead, seem to have  much more influence for Si/SiO$_2$ aggregates, as shown in  \cite{PRB79}. Hence, we have investigated the optical properties of the Si$_{32}$ and a-Si$_{32}$ embedded NCs including LF effects. The results are shown in Fig. \ref{LFvsNLF}. We note a strong similarity of the LF effects on the crystalline and amorphous systems with a progressive lowering of the spectra intensity as approaching the low energies. This matches well with the previous results that associates the low-energy absorption to the interface. Around the interface a strong charge discontinuity (inhomogeneity) occurs, and this could reasonably explain the strong LF effects in this region.

\section{Conclusions}\label{concl}
DFT calculations of the electronic and optical properties of a crystalline and an amorphous 32-atoms Si-NC embedded in a SiO$_2$ matrix are described and compared with that of the same NC extracted by the matrix with or without the interface oxygen atoms and passivated with H. We find that the NC/matrix heterojunction shows a type I character with the band edges states and consequently the low energy range optical spectrum is completely due to the Si NC plus the interface oxygens. Only higher in energy the response of the system is dominated by the matrix. This separation established between the NC+interface system and the silica matrix is also supported by the application of the Bruggeman Effective Medium Approximation that also underlines the importance of the strain induced by the lattice mismatch between Si and SiO$_2$ when the optical properties are calculated.
\\The optical absorption threshold for crystalline or amorphous embedded NCs has also been evaluated as a function of NC size finding that, beside the quantum confinement trend, the NC amorphization introduces an additional redshift due to which the band gap tend faster to the bulk limit.
\\Finally, the inclusion of local fields on the DFT-RPA optical absorption spectra of the crystalline and amorphous systems produces important redistributions of the transition probabilities in both cases, while not affecting the absorption thresholds. This supports the idea that, for a proper description of the optical response of the interface, inhomogeneity effects at the microscopic level (local fields effects) should be included.

\section{Acknowledgements}
This work was supported by PRIN2007, CNR Italia-Turchia and Fondazione Cassa di Risparmio di Modena. We acknowledge CINECA CPU time granted by INFM "Progetto Calcolo Parallelo" and CASPUR Standard HCP Grant 2009. We thank A. Pedone for the contribution with the modeling of the silica glasses. We acknowledge support from EU e-I3 ETSF project n. 211956.

\end{document}